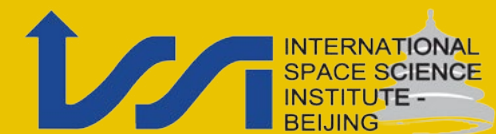

太空 | **TAIKONG**

**ISSI-BJ Magazine**

**No. 6 February 2015**

国际空间科学研究所北京

# Exploring the Dynamic X-ray Universe

**Front Cover**

This image artistically illustrates the Einstein Probe satellite and the process of a black hole tidal disruption event, in which a star is being tidally disrupted by a black hole close by and part of the stellar mass is accreted onto the black hole, giving rise to a tremendous flare of high-energy radiation mostly as soft X-ray and UV photons (image credit: NASA).

# FOREWORD

The International Space Science Institute in Beijing (ISSI-BJ) successfully organized a two day Forum on "Exploring the dynamic X-ray Universe" in the framework of the Space Science Strategic Pioneer Project of the Chinese Academy of Sciences (CAS), in Beijing on May 6-7, 2014. ISSI-BJ Forums are informal and free debates, brainstorming meeting, among some twenty-five high-level participants on open questions of scientific nature. In total 26 leading scientists from eight countries participated in this Forum coming from UK, France, Germany, Italy, Japan, Switzerland, USA, and China.

The Forum's main aims were to discuss the importance of the scientific use of soft X-ray wide-field monitoring observatories in the Violent Universe domain as well as to discuss the technologies. In the soft X-ray regime, the novel micro-pore lobster-eye optics provides a promising technology to realize, for the first time, focusing X-ray optics for wide-angle monitors to achieve a good combination of sensitivity and wide field of view. In this context, a soft X-ray all sky monitor called Einstein Probe is one of the candidate missions of the Intensive Preparative Study of Future Space Science Missions, which aims at selecting appropriate new space science missions to be implemented during the Five-Year Plan period 2016-2020. The Einstein Probe (EP) proposal should be submitted to CAS in 2015 to compete in the final selection.

The Forum started with an overview of X-ray all sky monitoring and missions. The participants examined also the prospects of detecting in X-rays the counterparts of gravitational wave events found with the next generation of gravitational wave detectors. The participants recognized the very high scientific value of the mission and raised constructive comments and suggestions for a better definition of the objectives and project organization. Therefore, the importance of this Forum has been recognized and appreciated by all the participants. In addition to all those very fruitful and stimulating discussions on the science and technology, it has made good international publicity and has in a way accelerated the project to a higher momentum.

The Forum participants concluded that the EP mission has complementary objectives to existing missions. Therefore, the EP mission is an additional excellent example of how the Chinese Space Science program is innovative and challenging, and complementary to existing missions. This offers significant opportunities for cooperation through mission coordination and scientific analysis that places EP and China in a central position due to its unique objectives and technology. This TAIKONG magazine provides an overview of the scientific objectives and the overall design of the EP project, including spacecraft and instrumentation discussed during the Forum.

I wish to thank the conveners and organizer of the Forum Weimin Yuan (NAOC, China), Julian Osborne (Leicester University, UK), Neil Gehrels (NASA/GSFC, US), George Fraser (†19th March, 2014, Leicester University, UK), Shuang-Nan Zhang (IHEP, China). I also wish to thank the ISSI-BJ staff, Ariane Bonnet, Lijuan En, and Xiaolong Dong, for actively und cheerfully supporting the organization of the Forum. In particular, I wish to thank Weimin Yuan and Julian Osborne, who with dedication, enthusiasm, and seriousness, conducted the whole Forum and the editing of this report. Let me also thank all those who participated actively in this stimulating Forum.

Prof. Dr. Maurizio Falanga
Beijing, February 2015

# INTRODUCTION

Transients and variable objects on timescales from sub-seconds to years pervade the X-ray universe, some as spectacular outbursts. Space-borne observations and theoretical studies of these violent displays have greatly advanced, even revolutionised, our understanding of the Universe and its underlying physical laws. Such tremendously high energy sources also provide ideal laboratories to explore the limits of contemporary physics and to study matter under extreme conditions. Exciting new phenomena still continue to be discovered which appeal for large-scale observations in order to characterise their natures. Other as-yet undetected classes of event are strongly predicted and presumably await discovery.

The previous two decades saw the opening up of the entire electromagnetic spectrum in not just mapping the sky but also studying at all wavelengths objects that we didn't properly understand before taking that multi-wavelength perspective. Over the next two decades we will now enter a period during which we investigate how the sky changes with time – time-domain astrophysics. This new age of discovery will also take place across the entire electromagnetic spectrum. Moreover, the advent of next-generation detectors such as ALIGO, AVirgo, and Ice-cube are extending our sight beyond only photons into the realms of gravitational waves and neutrinos.

The outstandingly capable missions and facilities that have already started to operate or are planned across the electromagnetic spectrum include LOFAR, the SKA, and LSST. We already have PTF and Pan-STARRS as particularly successful precursors to LSST, whilst at high energies Swift, MAXI and Fermi are operating in the X-ray and Gamma-ray bands. However, X-ray detection sensitivities are largely limited by the conventional non-focusing techniques. Micro-pore lobster-eye optics are a novel technology which promise to enable focusing X-ray optics with an excellent combination of sensitivity and wide field of view. A

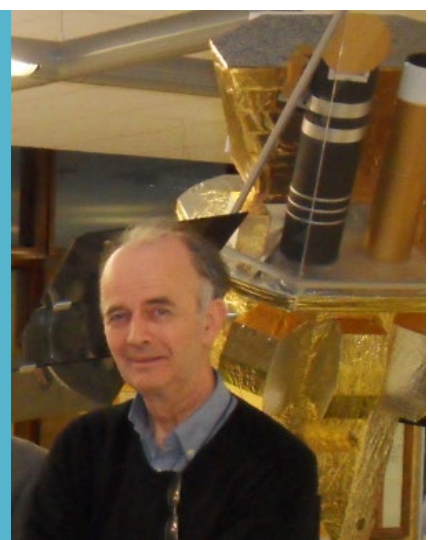

Fig. 1: Prof. George Fraser (1955-2014) (photo from W. Yuan).

*We wish to dedicate this forum as a commemoration of Prof. George Fraser, the late Director of the Space Research Center at Leicester University and one of the conveners of this meeting. George unfortunately passed away just before this forum. Among his major contributions to the technical development of X-ray astronomy, one is the innovative micro-pore lobster-eye technology applied to X-ray imaging. His work on X-ray optics is the basis of instruments proposed for several future space missions, including the Einstein Probe, and was one of the key topics discussed during this forum. George had kindly offered a great deal of advisory help for the building of the X-ray Imaging Lab at NAOC, and was an enthusiastic promoter of the collaboration with China on the Einstein Probe project.*

few mission concepts exist which take advantage of the capabilities of this emerging technology, including Lobster (proposed to NASA) and the Einstein Probe (EP) – proposed to the Chinese Academy of Sciences and now a candidate mission in the advanced study phase.

To discuss the driving scientific questions and new trends in this exciting field around the year 2020 and beyond, as well as how to achieve them technically, a forum was organized by ISSI-Beijing with the support from the NAOC on May 6-7, 2014. Some 26 participants from the UK, US, Italy, France, Germany, Switzerland, Japan and China were invited to meet together at ISSI-BJ located at the National Space Science Center (NSSC) for this lively 2-day brainstorming forum. We are honoured to have this high-profile group of international participants representing the state-of-the-art science and technology in this field, including PIs or project scientists of several current, previous and future missions (e.g. Swift, MAXI, BeppoSAX, LOFT, SVOM, Polar, Chinese SZ-2 experiment), renowned experts in space science, observational and theoretical astrophysics as well as space instrumentation. Special attention was given to the Einstein Probe, its key science goals and mission definition, as well as some of the key technical issues. With the enthusiasm and energetic support of all the participants, the forum was a great success.

# CHINESE SPACE SCIENCE PROGRAMME

The Director of the National Space Science Centre of CAS, Prof. Ji Wu, gave a brief introduction to the space science programme in make further contributions to the advancement of knowledge for the benefit of human beings. With the support of the central government, the Chinese Academy of Sciences (CAS) is now taking the lead in planning and developing China's space science programme. The forthcoming several years will outputs can be delivered by the selected missions, the mission science and feasibility should be discussed and scrutinized on an international platform. Thanks to the newly established ISSI-BJ, the space science forums it organizes provide a great opportunity to set up such an international stage.

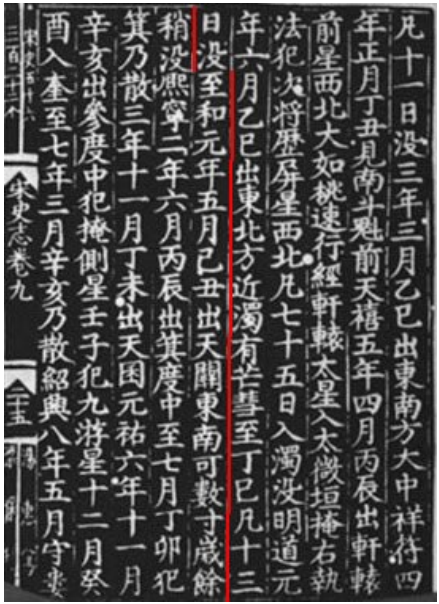 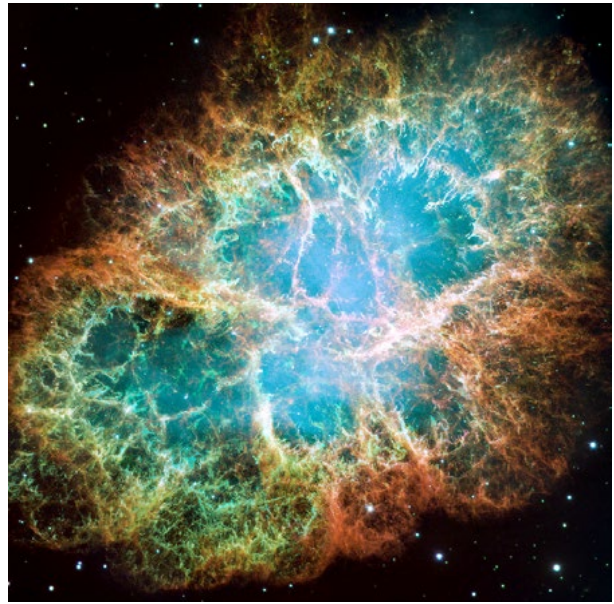

Fig. 2: "History of Song" documented the observation of a 'guest star' in the constellation Taurus on July 4, 1054 in the Chinese Song Dynasty (left, credit: Encyclopedia of China, 1999), which was later known to be a supernova explosion leading to the well-known leftover remnant known as the Crab Nebula (right, credit: Hubble Space Telescope).

China. China has a long history in the quest for the Universe. Back to a thousand years ago ancient Chinese had already observed and recorded some heavenly phenomena such as comets and supernova explosions. In the contemporary era, although China had already launched its first satellite in 1970, its space science programme had been lagging behind for decades. The first Chinese science mission was Double Star, launched in 2004, which used two satellites to investigate the Earth's magnetosphere in collaboration with ESA. In recent years China has achieved dramatic development in economics as well as in space technology, and China should see the launch of several approved science missions, including the Hard X-ray Modulation Telescope, Quantum Experiment on Space Scale, Dark Matter Explorer and Shijian-10 – a recoverable mission for microgravity experiments. To go further into the future, we need to plan future missions well in advance. Aiming for launch around the end of the next five-year plan starting from 2016, we have selected and supported eight missions for advanced study. These are called 'background missions'. Einstein Probe is one of them. There will be a further selection at the end of 2015 which will formally approve of a number of these missions. To ensure that cutting-edge scientific

# ALL-SKY MONITORING MISSIONS

Neil Gehrels reviewed the history and the present state of the art of all-sky monitoring in X-rays, as well as projecting into the likely future. Masaru Matsuoka reviewed the highlights of the MAXI mission – the most sensitive X-ray all-sky monitor so far.

**Previous missions**

Cosmic X-ray transients of diverse origins have been discovered since the early days of X-ray astronomy, thanks to successive all-sky monitors (ASM) or wide-field instruments operating in space in the X-ray and gamma-ray bands. The monitoring of the X-ray sky started with Vela 5B and Ariel 5. Vela 5B was the very first all-sky monitor that flew. Equipped with collimated X-ray detectors with a 6-degree by 6-degree field of view, Vela 5B scanned the whole sky every 56 hours. It is interesting that, looking back at the earlier papers, it was already very apparent in the late 1960s and early 1970s that Cygnus X-1 and other highly randomly-variable X-ray sources had something to do with black holes. Cygnus X-1 was already being talked about as a black hole candidate, and Cygnus X-3 was later discovered to be a micro-quasar, which we also believe to contain a black hole as well as producing jets.

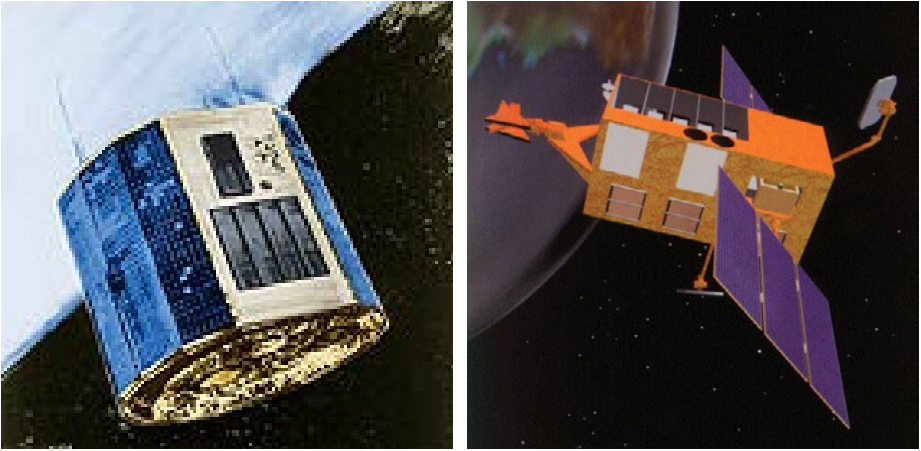

Fig. 3: Artist's impression of the Ariel-V (left) and RXTE (right) satellites (credit: NASA).

Ariel-5 was a joint British and US mission which had several instruments all doing time-domain astronomy. Its ASM adopted the interesting technology of a pinhole camera. This had a small aperture and a large detector area by the standards of the time, and so was sensitive enough to do all sky monitoring on a fast cadence. It was Ariel 5 that discovered the beautiful X-ray nova A0620-00, still the brightest nova ever detected. Such bright novae are rare and much can be learned from them.

One of the most successful ASMs was the ASM onboard NASA's RXTE, as it operated for a great many years and was extremely productive. It used an interesting design of three different scanning shadow cameras, essentially one-dimensional coded apertures, with a 6deg by 90deg FoV. It observed many X-ray flares of various types from X-ray binaries containing a neutron star or a black hole. These are similar in some ways to Cygnus X-1 but each system also had seemingly unique characteristics.

The power of supplementing a narrow-field, sensitive and fast pointing telescope to a wide-field monitor was demonstrated by BeppoSAX, an Italian-Dutch satellite operating in the late 1990s. By performing fast follow-up X-ray observations to study and precisely locate their afterglow X-ray emission, the cosmic origin of GRBs was established, with support from ground-based optical follow-up observations. This observational strategy is further highlighted by Swift, with the extraordinary success of its X-Ray Telescope.

## Missions in operation

NASA's Swift/BAT possesses the widest field-of-view (1.4 sr) and largest collecting-area detector (5200 $cm^2$) working in a hard X-ray band which is optimized for GRB detection. As the most powerful instrument for GRB detection Swift/BAT has detected almost 900 GRBs so far. The average number of BAT triggers is 3.3 per week, of which 1.8 are GRBs, 1.0 other transients, and 0.5 noise, which indicates that the current trigger threshold is well-chosen. Being a coded aperture instrument, BAT does imaging on the fly, which enables it to lower the trigger threshold to produce so many sensitive detections of GRBs. BAT has a daily sensitivity of 2-3mCrab, and so far the accumulated sensitivity has gone down to 0.3mCrab, which is good for long-term survey projects such as AGN detection. Swift has demonstrated that long and short GRBs have different origins, and has detected GRBs beyond a redshift of 8.

Swift also has a narrow-field focusing X-ray Telescope (XRT) and UV/Optical Telescope (UVOT), which are used to point to targets triggered by BAT or by other facilities. Swift/XRT is also performing ToO observations, on average three per day, for different kinds of transients from a large community, from supernovae to blazars and optical transients.

In addition to GRBs, Swift has also detected diverse and fascinating types of X-ray transient, including supernova shock breakout, tidal disruption events with relativistic jets, X-ray super-flares of young stars, super-bursts of thermonuclear Carbon burning from a neutron star in a low-mass X-ray binary, X-ray flares from the black hole at the centre of our Galaxy, etc.

MAXI (JAXA/RIKEN) is a Japanese experiment onboard the International Space Station. This is the most sensitive existing ASM, with a large detector collecting area of 5350 $cm^2$, pushing sensitivity to a new level. Employing slit aperture optics with collimated fan-like FoVs, it scans almost the whole sky once every 92 minutes. MAXI is the most prolific instrument in finding new transients and monitoring the sky sensitively in X-rays. After three years of operation MAXI has detected 500 X-ray sources brighter than 0.4mCrab (>7 sigma), ~40% of which have not been identified. MAXI has also discovered new 6 new black hole candidates and 6 new neutron stars. MAXI has detected and triggered a large number of transients and outburst events (123 ATel issues as of April 2014), of which 80% are Galactic and 20% are extragalactic or unknown events. These include X-ray novae (e.g. first discovery of a new type of X-ray nova thought to originate from the fireball phase in a very massive white dwarf), GRBs, an unbiased survey of stellar flares, X-ray transients in binary systems, TDEs, etc. Long-term X-ray light curves and spectra for a large number of BH/NS X-ray binaries, cataclysmic variables, AGNs, etc. have been obtained and have enabled further studies of detailed physics in compact systems.

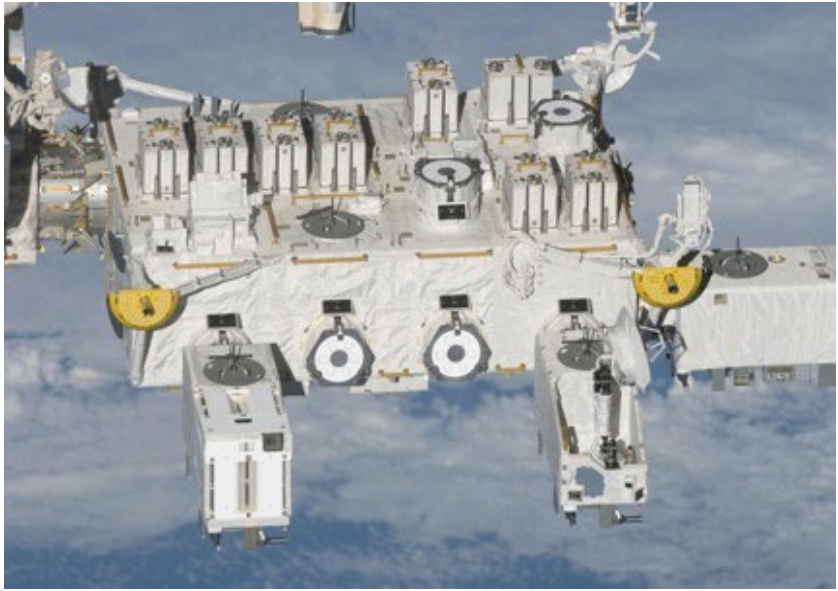

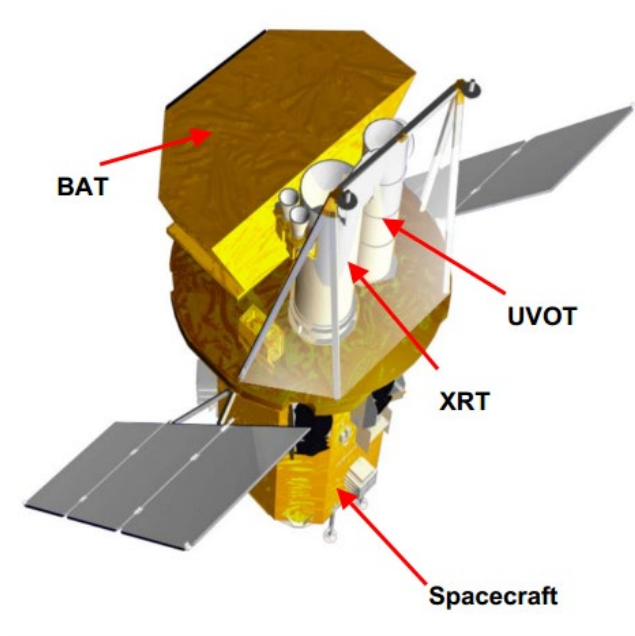

Fig. 4: MAXI (lower left) on the Exposed Facility of the International Space Station (top) and the Swift Satellite (below) (credit: JAXA/MAXI-team/M. Matsuoka; NASA/GSFC/N.Gehrels).

**Projection into the future: new science questions**

Recent discoveries by Swift, MAXI, and other wide-field instruments have presented new science opportunities in this field for the future, which are represented by the following four new directions.
1) Tidal disruption events: Do all events have jets? What is the signature of different types of stars? What is the physics of the accretion involved?
2) Supernova shock breakout: Do all SNe have X-ray shock-break outbursts? What is the signature of jetted vs. non-jetted SNe? Can SN explosions seconds after core collapse be detected by EP/Lobster?
3) High-redshift GRBs and gravitational waves: What was the nature of the first stars? When and how was the universe re-ionised? Are NS-NS mergers the origin of short GRBs?
4) Faint flares from compact objects: Can new systems be found to measure the mass and spin of the black holes and the equation of state of neutron stars?

**Summary**

The X-ray sky is rich in transients of various types. Historical and current all-sky monitors have advanced our understanding of the Universe and its extreme physical processes by exploring the X-ray sky in time domain by pushing the sensitivity from the about 100mCrab level down to about 1mCrab. Looking into the future the new technology of wide-field focusing X-ray optics will improve sensitivity by more than one or two orders of magnitudes, and will open up a wealth of new science.

# EXTRAGALAGTIC ALL-SKY MONITORINGt SCIENCE OPPORTUNITIES

**Tidal disruption events**

One of the key questions in astrophysics and cosmology is whether there is a massive black hole lurking at the centre of every single galaxy. Even as early as in the 1970's and 80's, the theory of tidal disruption of stars by massive black holes had already been conceived by a few pioneering theoreticians, in which an enormous accretion flare in X-rays was predicted. This, as pointed out by Martin Rees and the others, would be perhaps the most unique signature of the existence of massive black holes in the cores of otherwise quiescent galactic nuclei. Thus the detection of such tidal disruption events (TDEs) is of great interest in studies of the formation and evolution of black holes and galaxies in light of their co-evolution. Clearly it is also fascinating to study how matter falls onto black holes, and what happens when it does so.

A star that wanders too close to a massive black hole will be distorted and eventually disrupted once the formidable tidal force of the hole exceeds the self-gravity of the star, which takes place at the so-called tidal radius. A significant fraction of the stellar debris is expected to be accreted onto the black hole at a high rate close to the Eddington rate, producing an enormous flare of radiation with energies peaking at the soft X-rays and EUV. The mass return rate, as the stellar debris stretching across a number of different Keplerian orbits starts to fall back to the black hole, scales with time to the -5/3 power. This gives rise to the characteristic light-curves of TDEs as a rapid rise followed by a power-law decay, with timescales ranging from months to years. The event rate is estimated to be low, an order of one event per galaxy per ten to hundred thousand years. So, large field-of-view sky surveys in the soft X-ray band are best-suited to find such events.

As discussed by Stefanie Komossa, TDEs were first discovered in X-rays with the ROSAT mission in the 1990s, in the all-sky survey. Only a few events were identified. There is now increasing effort in the community to search for TDEs, mainly using serendipitous or systematic multiwavelength surveys. The most efficient waveband is X-ray, in which 14 TDE candidates have been currently found.

The remarkable discovery of Sw 1644+57 with the Swift satellite as a surprising TDE, from which a relativistic jet was launched, generated new excitement in TDE research, as reported by Paul O'Brien. Such intriguing yet rare events would shed light on the key question of how relativistic jets are formed and launched by massive black holes. In addition, Swift has found a small number of ultra-long (>2000s) GRB-like transients, whose relation with relativistic TDEs remains a mysterious question, which can be addressed by future missions such as Einstein Probe and Lobster.

As shown by Fukun Liu, if a galaxy has two massive black holes at its centre (binary black holes), the perturbation of the secondary black hole will lead to some distinctive features in TDE observations. Any TDEs produced by binary black holes should have a much enhanced event rate and distinctive drops and recurrences in their lightcurves. The latter might have been observed in at least one case, rendering TDEs as a useful tool to identify and study binary massive black holes.

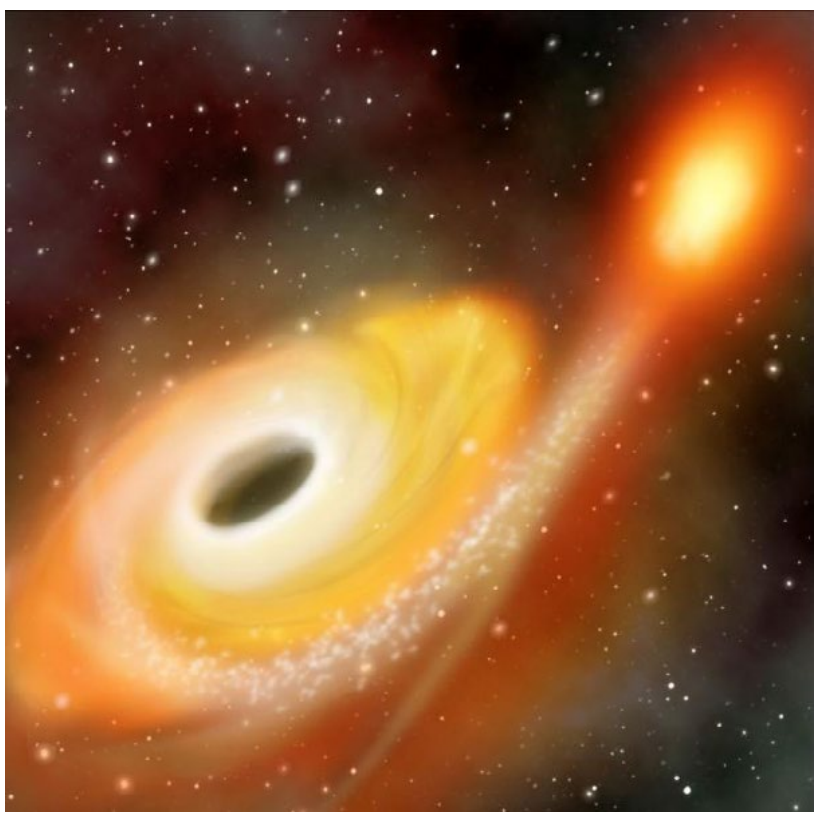

Fig. 5: Artist's impression of a stellar tidal disruption event by a massive black hole (credit: NASA).

**High-redshift GRBs as beacons to probe first stars and black holes and dark universe**

Luigi Piro reviewed the science prospects and requirement for the detection of high-redshift Gamma-ray bursts (GRBs), which are the most distant luminous objects in the Universe. The highest GRB redshift identified so far is 8.2, and it is expected that future transient finders should push this beyond redshift 15 (~2% of the age of the current Universe). Being the brightest objects (their X-ray afterglows outshine the most luminous quasars by orders of magnitude for one day), GRBs exploded, and thereby polluted their environment. The detection of these stars individually is impossible, however, even with the next generation of space observatories like JWST. The only way of observing Pop-III stars in action is to detect their explosive deaths. They have been predicted to produce a GRB-like event. By using high-redshift GRBs as beacons we could identify and probe the regions where the first stars and their remnants — the first black holes — were formed.

The first generation stars are often predicted to be typically of 50 solar masses or even higher. class in optical/infrared and X-ray observatories like Athena. In particular, high-resolution X-ray spectroscopy will be able to tell whether the GRB progenitor is in Pop-III or Pop-II environment. The candidates can be selected using the X-ray properties and supplementary information of possible photometric redshifts.

To detect these potential Pop-III collapsars, which are thought to be faint and long, would require sensitivity of <1mCrab and observing durations up to 10,000 seconds. This sensitivity is beyond the reach of the current and approved missions so far, and can therefore be used as powerful probes of the distant early Universe. They act as beacons shining through the cosmological dark ages.

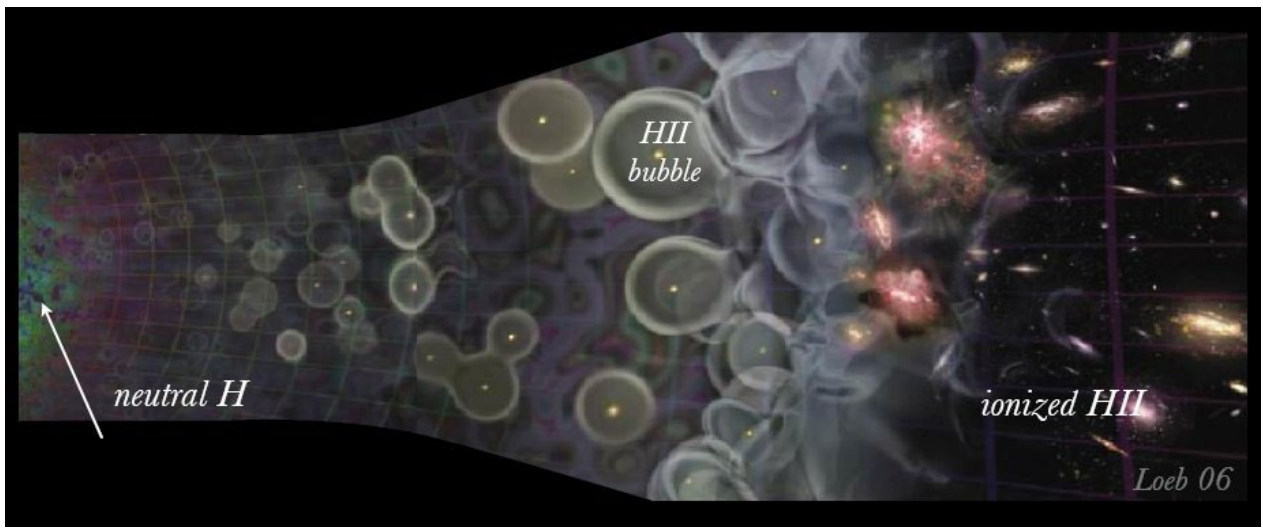

Fig. 6: Illustration of the dark Universe being re-ionised by the first generation (Pop-III) and later (Pop-II) stars at redshifts between 20 and 8. Gamma-ray bursts produced as the stars come to the end of their lives carry unique information about these stars and their environment (credit: A. Loeb 2006, Scientific American).

The early Universe between redshifts 20 and 6 is an exciting period, during which it is thought that the first luminous objects formed and started to re-ionise the Universe out of the dark ages. These are the so-called Pop-III stars, formed by clouds collapsing onto dark matter in a pristine environment. Whilst synthesizing metals they quickly evolved and Theoretical models predict that GRBs produced by such stars would have characteristic prolonged X-ray afterglow emission lasting up to 10,000 seconds, significantly longer than ordinary GRBs. In addition, GRBs produced by normal Pop-II stars, which could have formed immediately following the first Pop-III stars as early as at redshifts 15-20, will still carry information about the chemical abundance, dust, and molecular content in the early Universe. These will be the targets of the next generation of large telescopes of the 30-m can only be achieved with the X-ray focusing optics provided by the micro-pore Lobster-eye technology. A field-of-view as large as 1 steradian or more is also required given the low intrinsic event rates (~50 GRBs @z>7 per year). Given these expectations, the Einstein Probe should detect ~10 GRBs at redshifts beyond 7 over three years of operation, with a mix of Pop-II and Pop-III progenitors. This number should be increased when including the likely detection of orphan afterglows whose prompt burst emission is missed.

In conclusion, the identification of the site of formation of the first stars, black holes, and the first metals in the Universe will revolutionize our understanding of the early Universe. Even a few detections at truly high redshifts would be extremely rewarding. Einstein Probe will lead the way of time-domain observations of world's future large facilities by providing them with input targets.

**Supernova shock breakout**

Xuefeng Wu discussed supernova Shock Breakouts (SBs). At the end of their lives, massive stars undergo a catastrophic and spectacular process with their cores collapsed to form a neutron star or black hole and their outer shells ejected. A small fraction of the enormous energy released in this process at the centre somehow drives a radiation-dominated shock to propagate outwards. As the shock wave breaks through the stellar surface, a short yet intense burst of radiation is released, peaking at the extreme UV and soft X-ray wavebands. The typical durations predicted are from hundreds to thousands of seconds, with predicted peak luminosities which are comparable to those of active galactic nuclei. The duration and temperature of SB emission are determined by the size of the progenitor star, with brighter, cooler and longer bursts for larger stars and dimmer, hotter and shorter ones for smaller stars. As such, observations of SBs can give an insight into the size and other properties of the progenitor stars, and be used to examine theories of supernova.

Supernova SBs are elusive and their observational evidence is very preliminary, however, given their relatively faint and short-lived peak emission. There are only a few tentative observational results reported so far, which mostly caught only the declining phase of the bursts. There is only one promising candidate detected serendipitously with the XRT onboard Swift, which is a narrow-field X-ray telescope with a large effective area. Another notable example was detected in the UV by GALEX, which showed both the radiative precursor of the SB and the signature of the adiabatic expansion of the stellar surface after the breakout of the shock, but which could have been much more powerful if a simultaneous X-ray light-curve had existed.

Obviously, the opening up of this research topic relies on the detection of more SB events with compelling evidence and complete sampling of the light-curves. This can be achieved with the high sensitivity and wide field-of-view enabled by the MPO technology, as is of Einstein Probe.

**Unusual Gamma-ray bursts**

The detected GRB population has revealed several subtypes with unusual properties that remain poorly understood, as discussed by Xuefeng Wu.

Low-luminosity GRBs: These mysterious Gamma-ray bursts were observed to have isotropic luminosities several orders of magnitude lower than the bulk of the population. Though only detected in a small number (~6) in the local Universe, taking into account their low luminosity suggests that they may outnumber typical GRBs by 2-3 orders of magnitude. Curiously, 4 of the 6 known events were also associated with a supernova explosion. This may suggest their intimacy with mildly relativistic supernova shock breakout. Apart from their possibly large space number density, they are potentially important targets of multi-messenger research since, given their high event rate, they are potential sources of gravitational wave and neutrinos in the local Universe. The Einstein Probe is expected to detect low-luminosity GRBs at a rate at least 10 times higher than the previous and existing missions.

X-ray flashes: First noted by Ginga/ASM and later established by Beppo-SAX and HETE-II, these are X-ray transients with their lightcurves resembling those of GRBs but without Gamma-ray emission. Their nature is unknown, though several models have been suggested, including high-redshift GRBs, GRBs viewed off-axis, baryon-choked GRBs, or a mix of them. Further testing of these models is hampered by the small number of detections made so far and a lack of redshift measurement in general. The Einstein Probe is ideal for detecting X-ray flashes and is expected to enlarge the sample size significantly, taking advantage of the fact that their radiation peaks in soft X-rays. The fast alert capability will allow redshift measurements of their host galaxies by multi-wavelength follow-up observations.

# GALACTIC ALL-SKY MONITORING SCIENCE OPPORTUNITIES

Phil Charles showed how NASA's Rossi Timing Explorer had revealed the wide variety of types of variability to be seen in Galactic X-ray sky, with the modest All-Sky Monitor instrument providing light curves showing orbital modulation, accretion-driven outbursts, and many other forms.

X-ray binaries mainly come in two major classes. The high mass X-ray binaries have early-type companions, and virtually all of them have neutron star as the accreting object. They may have Be or supergiant secondaries, around which the neutron star travels in a highly eccentric and long period orbit, resulting in periodic outbursts as the compact object passes through the denser region close to the secondary star, perhaps forming a transient accretion disk and eventually switching to pseudo-spherical accretion far from the companion.

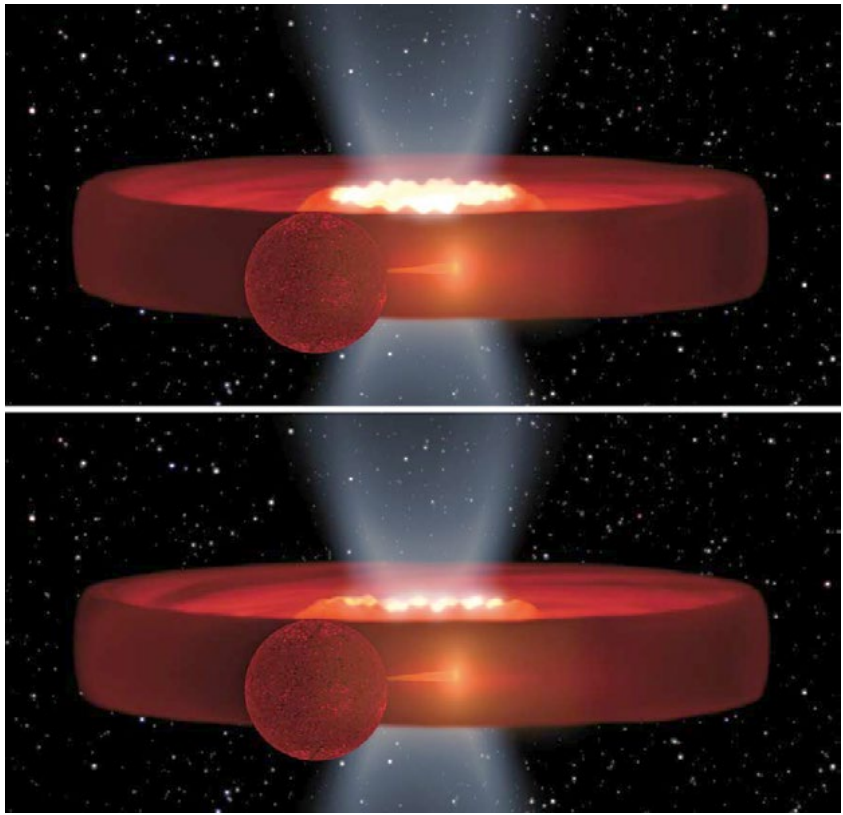

Fig. 7: The eclipsing low mass X-ray binary Swift J1357.2-0933 has an orbital period of 2.8 hours and an inclination of 85 degrees. It contains a black hole of more than 3 solar masses, but this is obscured by an inner disk structure that has a Keplerian rotation period of 3-8 minutes (Corral-Santana et al. 2013, Science 339, 1048). The disk at optical maximum is shown in the upper panel, and at minimum in the lower panel.

By contrast, the low mass X-ray binaries are short period systems that undergo rare and very bright outbursts, often separated by decades. Also known as Soft X-ray Transients or X-ray Novae, these systems can increase in optical brightness during outburst by up to 8 magnitudes due to strong irradiation of the accretion disk and companion star. In their outbursts, which may last more than a hundred days, they can produce strong radio jets, in which case they are known as micro-quasars; ~75% of them contain a black hole.

In general, the presence of bursts or pulsations reveals the presence of a neutron star; otherwise a black hole may be present. X-ray bursts may recur on hours to many days in low mass X-ray binaries, with almost half of them showing this behaviour; a few show very long 'super-bursts'. The normal bursts are due to the explosive ignition of accreted hydrogen on the neutron star, but the super-bursts are due to carbon ignition. Definitive identification of the nature of the compact object requires optical radial velocity measurements around the orbit, with rotational broadening to give the mass ratio and ellipsoidal flux variations to constrain the system inclination. However, only less than half of the suspected black hole X-ray binaries have had their mass dynamically determined. Einstein Probe can help by discovering more black hole systems so that the true distribution of BH masses can be measured, and to see if there is a continuous distribution down to the neutron star mass limit.

The recent discovery of the first eclipsing black hole low mass X-ray binary, Swift J1357.2-0933, among ~50 Galactic black hole transients shows us for the first time evidence of the suspected hidden high-inclination systems. Optical flux variations at frequencies much greater than that of the orbital period imply obscuration by the structure of the inner accretion disk rather than the previously suspected disk rim. Such obscured systems may be associated with the otherwise mysterious very faint X-ray transients.

The reprocessing of high-energy radiation in X-ray binaries in outburst leads to optical line emission, which is key to determining their structure. Narrow emission lines from the irradiated companion can be used to measure the mass function in low mass X-ray binaries.

Low mass X-ray binaries can show radio emission from jets in their hard states, with multiple ejections occurring at transitions to high states, e.g. in the micro-quasar GRS 1915+105. Einstein Probe can trigger the key radio observations to help us understand these structural changes in the accretion around black holes.

High mass X-ray binaries containing Be stars often show transient X-ray emission on the orbital period as the neutron star encounters the equatorial disk of the companion on its wide and eccentric orbit. They can also exhibit modulation at the neutron star spin period, and also at super-orbital periods of years or more; this latter modulation reflecting the timescale of formation and dispersal of the Be star disk.

The Corbet diagram is a useful means of classifying high mass-X-ray binaries. In a plot of spin period against orbital period the various systems resolve into islands of specific types, with wind- and Roche-lobe-fed supergiants having shorter orbital periods, while the Be X-ray transients have periods above 10 days.

The Small Magellanic Cloud has a remarkable number of X-ray pulsars, over 65 are known. This is a valuable population, as they are all at a common, known distance and exist in similar environments. There is a bimodal distribution of spin periods in these Be systems, which may reflect the two formation mechanisms, iron core collapse supernovae and electron capture supernovae. More such systems are needed to firm up on the distributions of system parameters in order to verify this proposed formation diagnostic.

In contrast to the Be X-ray binaries, the supergiant systems are persistent X-ray sources with circular orbits. They are characterised by rapid flaring, indicative of a structured wind from the supergiant. Some of these systems are heavily obscured by their wind, and perhaps represent the common-envelope spiral-in phase that follows after the high mass X-ray binary stage in their evolution.

A recent surprising discovery has been the detection of >100 MeV gamma-ray emission from classical novae by the Fermi-LAT. The first of these was V407 Cyg, a system in which the ejecta from the white dwarf eruption expand in the dense wind of the red giant binary companion. It was the shock of the ejecta with this wind that was initially thought to be the mechanism of the gamma-ray emission, but subsequent classical novae occurring in shorter period systems ruled out this hypothesis, as the secondaries could not be massive enough to drive the dense wind required.

Galactic binary X-ray sources show a wide variety of periodic behaviours, offering the possibility of full system diagnosis if optical spectroscopy can be used to disentangle their structure. In general, multi-wavelength observations are key to understanding them, and in the coming years there will be a wealth of facilities with which to achieve this. LCOGT, CTS, LOFAR are already operational, and we look forward to the full realisation of Z-PTF, PanSTARRS, GAIA and LSST (just to name some) in order to achieve this. However, all these facilities need wide-field X-ray facilities to discover the X-ray binaries and their outbursts; Einstein Probe is the ideal instrument to open up the new era of time domain astrophysics.

Xiangdong Li highlighted the potential of Einstein Probe to contribute to the understanding of the various types of Galactic transients and variable sources. He described the case of the supergiant fast X-ray transients, which display flaring activity on few-hour timescales with extreme dynamic range (up to 5 orders of magnitude). These systems appear to be accreting X-ray pulsars, but this recently recognised class of ~10 confirmed members are poorly understood: What is the link between them and the persistent supergiant X-ray binaries? How do they form? Are their multiple evolutionary histories? Are the companion star winds much clumpier than those of the persistent sources? Are the orbits systematically different? Do magnetic or centrifugal barriers play a role in the extraordinary variability seen? Einstein Probe can obtain a full census of the Galactic population; measure their true duty cycle; measure their periodicities; and model their broad-band spectra. Perhaps it could determine whether magnetars are present in these systems.

The outbursts of the Be X-ray binaries have a peak luminosity that correlates with the neutron star spin period. It is not clear whether this reflects different accretion modes in these systems. At lower luminosities advective or quasi-spherical accretion is inefficient at spinning up the neutron star, whereas in the giant outbursts a radiatively cooling disk is formed which is much more efficient. Einstein Probe should look for new types of Be X-ray binary, such as Be+WD and Be+BH binaries, for which very few examples are currently known.

The formation of black holes in low mass X-ray binaries is not well understood, with over a thousand inferred from observations to be present in our Galaxy compared to the predicted formation rate which suggests there should be less than ~100. The discovery of BH LMXBs at high Galactic latitude (implying high formation kick velocities), and in Globular Clusters (where simulations give contradictory results) will inform discussion of their formation mechanism.

There remains much to be learnt about the outbursts of X-ray binaries. Complex changes in the accretion disk structure, with disk evaporation, coronal condensation and the formation of jets all inferred. Observations of the luminosity at state transitions are vital to make further progress.

The very faint X-ray transients, found in large numbers close to the Galactic centre, have a peak luminosity several orders of magnitude below typical transient values. A third show type 1 X-ray busts, so must contain neutron stars; however the low accretion rate cannot be reached by conventional evolutionary means. Accretion may be occurring from a brown dwarf or a planet, or perhaps an intermediate mass black hole is accreting from a very low mass star. They represent a new regime in which to study accretion in X-ray binaries, and are sure to expand our understanding of the evolution and outburst mechanisms that may exist in these systems.

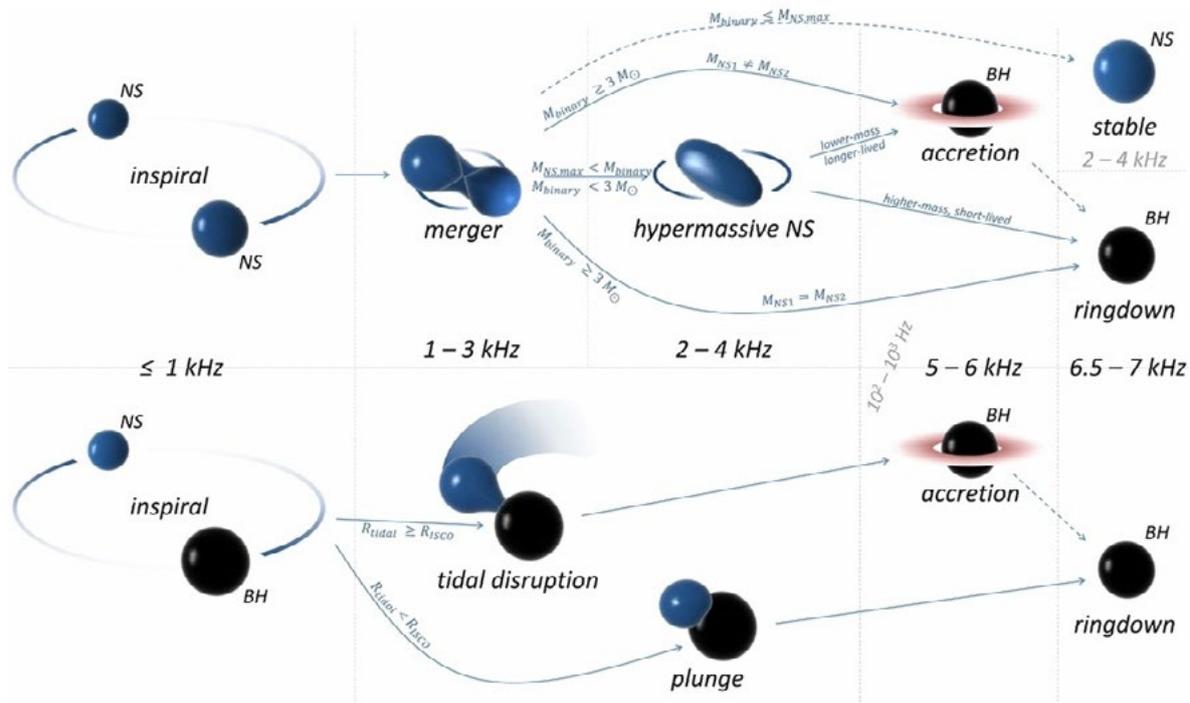

Fig. 8: Possible pathways of merging neutron star and neutron star - black hole binaries with the characteristic gravitational wave frequencies given for the various phases (credit: Bartos et al. 2013, CQG 30, 123001).

Einstein Probe will have a valuable combination of a very large field of view with high sensitivity. In addition to discovering new types of Galactic X-ray transient systems, it will greatly improve our understanding of the formation and evolution of black holes and neutron stars, the mechanisms of their outbursts and state transitions, and the physics of their accretions disks.

## SYNERGIES WITH MULTI-MESSENGER FACILITIES

A high-cadence X-ray sky survey can be expected to detect many new X-ray sources, and perhaps also new types of X-ray source. It will therefore be important to gather together all the available information on these new sources and their precursor objects promptly so as to maximise the likelihood of achieving a quick classification. Much information will come from observations at other wavelengths, but we are now reaching an era in which non-photon astrophysics is becoming possible, and we can expect this to play a major role in the coming decade.

Patrick Sutton, on behalf of the LIGO scientific and Virgo collaborations, described the future prospects for the detection of gravitational wave transients. Mergers of binary neutron stars or a neutron star with a black hole are expected to be strong transient gravitational wave sources with significant electromagnetic counterparts, with about one NS-NS merger per year within 100 Mpc. Advanced GW detectors are now being constructed by the LIGO and Virgo teams, aiming eventually for five interferometers around the world to reduce the location uncertainty of any detected sources. For example, Advanced LIGO is expected to commence operations in 2015, reaching full design sensitivity by 2019. Advanced LIGO and Virgo detection rates are rather uncertain, with estimates between 0.2 - 200 NS-NS detections per year by 2019; in addition the GW sources are hard to locate, such that only ~10-30% will have positions better than 20 square degrees. NS-NS mergers are thought to be the origin of the short GRBs, and simultaneous detection of an SGRB-like X-ray transient with a GW transient would provide the definitive confirmation of this model. Additionally, the GW signal can provide a measure of the system mass, BH spin, while joint GW-X-ray detection of a sample will tell us the opening angle of the highly relativistic jet that is the actual source of the gamma-ray emission. The teams are now preparing for the joint GW-photon transient era, with rapid alert systems and collaboration agreements being developed. The Einstein Probe has the potential to be a major contributor to this exciting new field.

The many possible mechanisms for generating significant photon fluxes from gravitational wave transients were emphasised by Bing Zhang. In the case of a NS-NS or NS-BH merger these fall into three broad classes: the collimated jet (causing a GRB if the jet points at us), a merger nova (sometimes called a kilonova), and the non-relativistic shock between the ejecta and the interstellar medium. While short GRBs are well known (if not well understood), there does appear to have been a first detection of a merger nova in GRB 130603B, in which a faint IR source was seen ~10 days after the burst, consistent with recent work from this peaks at soft X-ray energies, and has a detectable flux for up to 3 hours. This wind can also enhance the merger nova output. The ejecta-ISM shock can also have energy injected from this wind; the X-ray, optical and radio peak fluxes and timescales being diagnostic of the magnetic field and ejecta mass. The multi-wavelength light curves of the optical transient PTF11agg and some short GRBs are consistent with this model. Because isotropic emission is expected, it is important that early X-ray observations of GW transients are performed; these are best achieved with a large field of view high-cadence survey with rapid follow-up capability.

Julian Osborne described the actual follow-up of GW and neutrino triggers with the current of such electromagnetic follow-up to the GW trigger in terms of detection efficiency as a function of false alarm rate, with a ten-time increase in efficiency resulting from the Swift observations. These observations were a valuable means to learn how to maximise the value of such follow-ups, and subsequent improvements in Swift mosaicing, faint X-ray source detection and serendipitous source characterisation will all help in the Advanced LIGO/Virgo era. The improved X-ray source detection, which gives more than a factor two increase in detection efficiency at low fluxes with a much-reduced false positive rate, has subsequently been used to make the first fully comprehensive Swift XRT point source catalogue, 1SXPS.

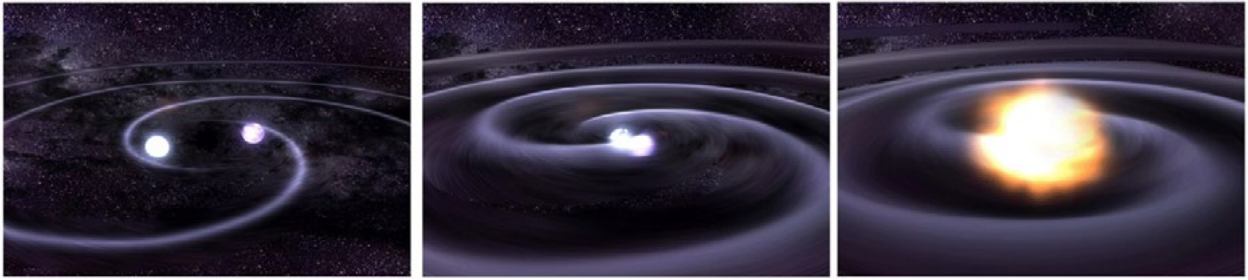

Fig. 9: Visualisation of a pair of neutron stars in the process of merging, with their strong gravitational wave emission propagating outwards at the speed of light (credit: NASA/CXC/GSFC/T. Strohmayer; http://www.ligo.org/science/GW-Inspiral.php).

suggesting a high heavy element opacity and thus a peak in the IR band. The uncertain presence of a highly massive magnetar as an intermediate merger product can have profound impact on the emission timescales, perhaps leading to a magnetic dissipation X-ray afterglow, which may be the origin of the plateaus seen in the X-ray light curves of some short GRBs. Such massive magnetars have been shown to form from NS binaries of modest mass. Initially rotating with a period of ~ 1 millisecond, these objects have a huge potential spin-down energy which can power an isotropic magnetar wind; Bing Zhang's own work has shown that the emission Swift narrow field X-ray and UVOT telescopes. The highly automated and manoeuvrable Swift satellite is able to make rapid mosaics of short pointings promptly in response to external alerts. In addition to the neutrino alerts from IceCube and ANTARES followed by Swift, two LIGO-Virgo alerts resulted in Swift observations in 2010. The first, in January, resulted from an artificial reduction in the alert generation threshold, while the second, in September, was artificially introduced as a system test; neither was thought to be a real GW event, and both had error regions widely spread across the sky. A sophisticated statistical analysis showed the added value The planned very-high-energy gamma-ray observatory, the Cherenkov Telescope Array, was shown to have superb capabilities for GRB and GWB follow-up, even if the anticipated numbers of detections is small. Follow-up of transient alerts is feasible with such traditional narrow-field high-energy telescopes, but a wide-field telescope is much more likely to catch the bright prompt phase of these events.

The capabilities of a new gamma-ray polarimeter, POLAR, to be flown on the Chinese Spacelab, Tian-Gong, in late 2015 were shown by Shuan-Nan Zhang to be highly valuable at energies over 50-500 keV. With an effective area of ~120 cm$^2$ and a field of view covering around one third of the sky, this Chinese/Swiss/Polish/French Compton scattering instrument is expected to be sensitive to 30% polarisation for 60 GRBs a year, and to 8% polarisation for 10 per year, presuming that independent position and spectral information is available.

Square pore MPO X-ray optics can be used in two geometries, the Wolter I (Wolter 1952) configuration as used in the conventional X-ray telescopes of Chandra and XMM-Newton Observatories or the lobster eye as described by Angel (1979). Grazing incidence reflection of X-rays from 2 surfaces is required to form an image in both cases. For the Wolter I these reflections occur in the same plane while for the lobster eye they occur in 2 perpendicular planes. The MIXS-T instrument on the ESA BepiColumbo mission uses tandem pairs of square pore MPO, in which the pores are packed in a radial pattern, to produce an approximation to the Wolter I geometry. The aperture and arrangement of the MPO of MIXS-T are shown in Figure 10.

The full width at half maximum (FWHM) of the MIXS-T flight model is 8.5x7.5 arc minutes which meets the requirements but it is limited by the quality of the radially packed plates, difficulties of aligning the 2 plates within each sector and the problem of aligning each tandem pair to produce a common focus.

The lobster eye geometry can be used for a narrow field or wide field X-ray telescope. Figure 11 right shows a schematic of the optical bench of a narrow field telescope with focal length ~1 m that might play the same role as the

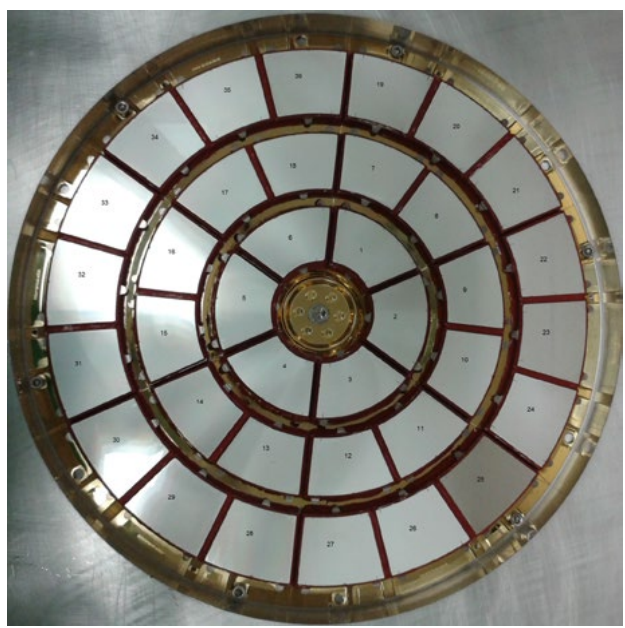
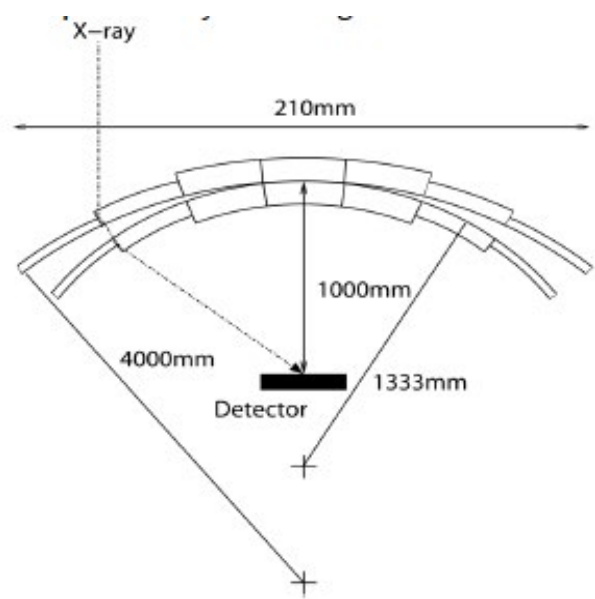

Fig. 10: Left: The MIXS-T aperture (credit: R. Willingale, the University of Leicester). Right: The curvature of the tandem plates (not to scale) (credit: Fraser et al. 2010 P&SS 58, 79).

## NOVEL MICRO-PORE OPTICS TECHNOLOGY FOR X-RAY FOCUSING IMAGING

Richard Willingale reviewed the principle and status of the micro-pore technology for X-ray imaging. The driving science of X-ray time-domain astronomy calls for high detecting sensitivity, good angular resolution (1 arc-minute or less) and all-sky coverage (field of view of order of 1000 square degrees for a module). Micro-pore optics (MPO), or Micro-Channel Plates as sometimes termed, for X-ray focusing can fulfil these requirements, as focusing results in enormously enhanced gain in signal to noise, and thus high detecting sensitivity.

Swift XRT in follow-up soft X-ray observations of bright transients. The aperture comprises an array of 21 square pore MPOs. Figure 11 left shows a wide field version with a focal length of 0.3 m and an aperture formed by an array of 7x7 square pore MPOs. In both cases the MPOs are 40x40 mm$^2$ and are mounted on a spherical frame with radius of curvature 2 times the focal length.

The point spread function of the lobster eye has a central spot produced by 2 reflections from adjacent sides of the pores and cross-arms produced by single reflections within the pores. A lobster eye telescope constructed using MPOs is much easier to align than the Wolter I equivalent and using the MIXS-T results we predict that the central spot will have a FWHM ~4.4 arcmins and focusing gain ~1900. If we include the efficiency of a typical CCD detector and light blocking filter the effective area of the central spot has a peak value of 32 cm$^2$ at ~1 keV for the narrow field and 1.5 cm$^2$ for the wide field version.

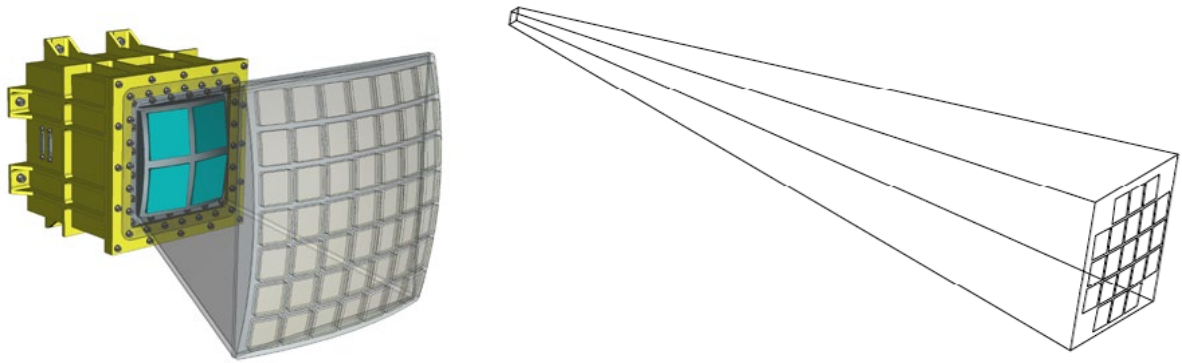

Fig. 11: Lobster eye telescopes. Left: wide field. Right: narrow field. (Credit: R. Willingale, University of Leicester)

The combination of an array of wide field lobster eye modules to act as a soft X-ray transient monitor and a narrow field lobster eye telescope for follow-up observations gives unprecedented grasp and sensitivity which will revolutionise X-ray transient astronomy.

# SCIENTIFIC OBJECTIVES OF EINSTEIN PROBE

Weimin Yuan, on behalf of the Einstein Probe team, presented the scientific objectives of the mission. The Einstein Probe will carry out a systematic census of the population of high-energy transients in soft X-ray with unprecedented sensitivity, over a wide range of timescales and at high cadence. It will also perform immediate follow-up observations with its onboard narrow-field telescope and will issue prompt alerts to trigger and guide follow-up observations by worldwide multi-wavelength facilities.

**Key scientific objectives**

The mission will focus on the following key scientific objectives:
1. Discover quiescent black holes at almost all astrophysical mass scales and study how matter falls onto them by detecting transient X-ray flares, particularly stars being tidally-disrupted by otherwise dormant massive black holes at galactic centres.
2. Discover and precisely locate the X-ray photonic counterparts of gravitational-wave events (such as binary neutron star mergers) detected with the next generation of gravitational-wave detectors.
3. Carry out systematic and sensitive surveys of high-energy transients, to discover faint X-ray transients, such as high-redshift GRBs, supernova shock breakout, and new types of transients.

In a sense EP is a black hole finder. In contrast to other previous and current X-ray missions with narrow field-of-view, which can only observe active black holes, EP will target non-active (quiescent) black holes – the main constituents of the black hole population in the Universe. It is hypothesized that massive black holes are present in the centre of almost every galaxy and the vast majority are lurking in a quiescent state and so remain currently undetected. EP will discover their presence by detecting the tidal disruption of stars (TDE) by these black holes. The high observing cadence enables such events to be detected at the onset of the process, which provide valuable information on the disruption process and how matter falls on to a black hole, and how relativistic jets are produced. EP is expected to discover from tens to hundreds of TDEs per year. EP will also discover new stellar- and intermediate-mass black holes lurking in our and nearby galaxies, by detecting their outbursts due to some kind of instability of gas accretion.

The detection of gravitational wave signal is perhaps one of the most important events in physics and astrophysics, and will close the last missing link of the prediction of General Relativity. By synergy with the next generation of GW detectors EP has a great potential to detect the photonic counterparts of GW events, which are essential for locating them and identifying their astrophysical origin. The detection of both, GW and their associated photonic sources, is the basis of gravitational-wave astronomy, whose advent is highly anticipated in the near future.

The first generation of stars (Pop-III) and their early descendants (Pop-II stars) in the early Universe between redshifts 20-6 are almost undetectable individually, except for the gamma-ray bursts potentially produced at the end of their lives. EP is expected to detect such GRBs at redshifts beyond 10 at a rate of a few per year. Follow-up observations of their afterglows with the EP narrow-field follow-up telescope in X-ray and other next-generation multi-wavelength facilities worldwide will shed light on the physical property of their progenitors and environment, as well as the re-ionisation in the early Universe.

We also expect that EP would regularly observe the predicted but elusive soft X-ray bursts which are generated as outward-propagating shocks break out of the stellar surface, following core collapse of massive stars in the making of supernovae. The X-ray lightcurves and spectra of such events hold the key to probe some of the parameters of the progenitor stars and the process of supernova explosion. Seeing shock breakout events is tightly linked to the detection of neutrinos directly from core-collapse. Given their relative faintness and short timescales, only sensitive monitors working in the soft X-ray band with large field-of-view like the Einstein Probe can perform systematic surveys for such shock breakout events.

Other known types of high energy transients to be detected with Einstein Probe include, but not limited to:

- X-ray flashes
- Low-luminosity GRBs, X-ray rich GRBs, normal GRBs and GRB precursors
- Soft gamma-ray repeaters (magnetars)
- Stellar corona flares
- Classical novae
- Supergiant fast X-ray transients
- Outbursts of active galactic nuclei and blazars

As by-products, Einstein Probe will monitor the variability of large samples of various types of known X-ray sources all over the sky over a wide range of timescales. It will also produce an all-sky image of the soft X-ray sky at an unprecedented depth.

## Scientific impact

The mission will address questions such as the prevalence of black holes in the Universe, how black holes accrete mass and launch jets, the first generation of stars and re-ionisation in the early Universe, the astrophysical origins and underlying processes of gravitational wave events, and the physics which operates in extreme conditions of strong gravity. The scientific impact of Einstein Probe will span a vast range of astronomy and astrophysics research, from comets, stars, compact objects in our and nearby galaxies, black holes, supernovae, galaxies to cosmology, as well as fundamental physics.

# OVERVIEW OF THE EINSTEIN PROBE MISSION

Weimin Yuan presented an overview of the Einstein Probe mission, including the scientific requirements, conceptual design and scientific capabilities. Chen Zhang and Hua Feng discussed the design and the status of development of the X-ray optics and focal plane detectors, respectively.

The proposed Einstein Probe mission is a small satellite to discover and characterise high-energy transients and monitor variable objects in the X-ray band (0.5-4 keV). The monitoring instrument is a Wide-field X-ray Telescope (WXT) with a large instantaneous field-of-view (60°×60°, 1.1 steradian). The wide-field imaging capability of WXT is achieved by using the micro-pore lobster-eye optics, thereby offering unprecedentedly high sensitivity and large grasp superseding previous and existing X-ray all-sky monitors and survey missions by orders of magnitude. To complement this powerful ability to monitor and discover sources over a wide area, the Einstein Probe will also carry a smaller field-of-view (1°×1°) Follow-up X-ray telescope (FXT) — capable of much larger light-collecting power and better energy resolution than the main survey telescope — with which to perform follow-up observations of newly-discovered transients. Public transient alerts from EP will also trigger multi-wavelength follow-up observations from the worldwide community. The proposed launching date is around 2020 and the nominal mission will last for three years (with a goal of five years).

|  | WXT | FXT |
| --- | --- | --- |
| Field-of-view | 60°×60° | 1°×1° |
| Focal length | 375 mm | 1400 mm |
| Energy band | 0.5-4 keV | 0.5-4 keV |
| Effective area | 3 cm$^2$ (@0.7 keV, central focal spot) | 60 cm$^2$ (@1 keV) |
| Angular resolution | <5' | <5' |
| Sensitivity (@1ks) | About 1x10$^{-11}$ erg/s/cm$^2$ | About 3x10$^{-12}$ erg/s/cm$^2$ |
| Timing resolution | 100 μs | 1 s |
| Energy resolution | ~50% @ 4keV | ~100 eV @ 1keV |

Table 1: Specifications of WXT and FXT

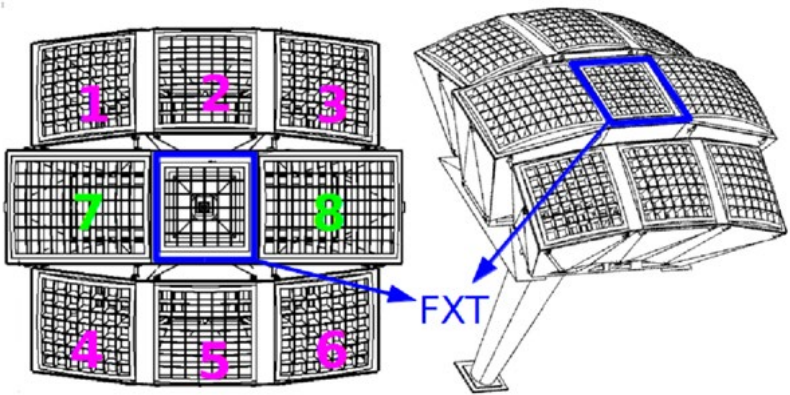

Fig. 12: Layout of WXT and FXT (courtesy of NAOC).

The WXT consists of eight modules, with the FXT mounted at the centre. Both the WXT and FXT optics are being developed at the X-ray Imaging Laboratory at NAOC (PI: Chen Zhang). WXT employs the MPO lobster-eye optics, with a focal length of 375 mm. The eight WXT modules make up a spherical array mosaicked by 434 confocal MPO pieces, each of 40mmx40mm in size. The total FoV subtends a solid angle of 60°×60° (1.1 steradian), which is about 1/11 of the whole sky. WXT has a large focal plane size, 420 mm*420 mm in total. The baseline choice of the WXT detector is gas detector, which has been the workhorse in X-ray astronomy for decades. Currently, the employment of gas detectors is being investigated and the GEM-based (Gas Electron Multiplier) detectors are being developed at Tsinghua University (PI: Hua Feng). As a back-up plan, microchannel-plate (MCP) detectors, which have the benefit of being a mature technology and widely used in X-ray astronomy, are also considered as a surrogate for the WXT focal plane detectors.

The FXT is a narrow field-of-view telescope of the MPO type, with a focal length of 1.4m, leading to a much larger effective area (~60 cm$^2$ @1 keV). The FXT has an aperture size of about 240 mm, which is mosaicked by 6*6 MPO pieces. The focal plane detector of FXT will employ silicon-based type detectors (i.e. CCD) to gain better spectral performance.

For the fast telemetry of alert data, a promising approach is to make use of the VHF system by collaboration with the French CNES. An alternative is to make use of the Chinese relay satellites network.

Scientific capability of EP:
- Monitoring a large fraction of the whole sky at high cadences with X-ray sensitivity at least one order of magnitude deeper than the most sensitive wide-field monitoring instruments (e.g. MAXI, Swift/BAT).
- Detection and classification of transients by real-time processing of WXT data on-board.
- Triggering prompt follow-up observations in the X-ray band to measure the position, spectra and variability of the transients at a much better precision.
- Prompt downlink of the alert data of transients and broadcast them to global observing facilities
- Pointed observations of targets of opportunity triggered by other facilities.

As a comparison, the detecting sensitivity of Einstein Probe is deeper than Swift/BAT by about two orders of magnitude, and than MAXI by more than one order of magnitude. A measure of the survey capability of an instrument is grasp – the product of effective area and field-of-view. The Grasp of EP/WXT is the largest among the focusing X-ray instruments of the past, current and planned future missions.

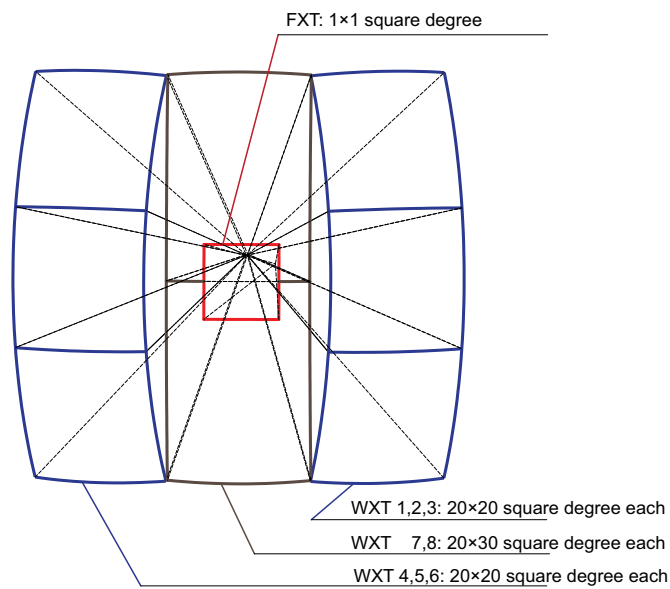

Fig. 13: Illustration of the field-of-view coverage of WXT and FXT modules (not to scale).

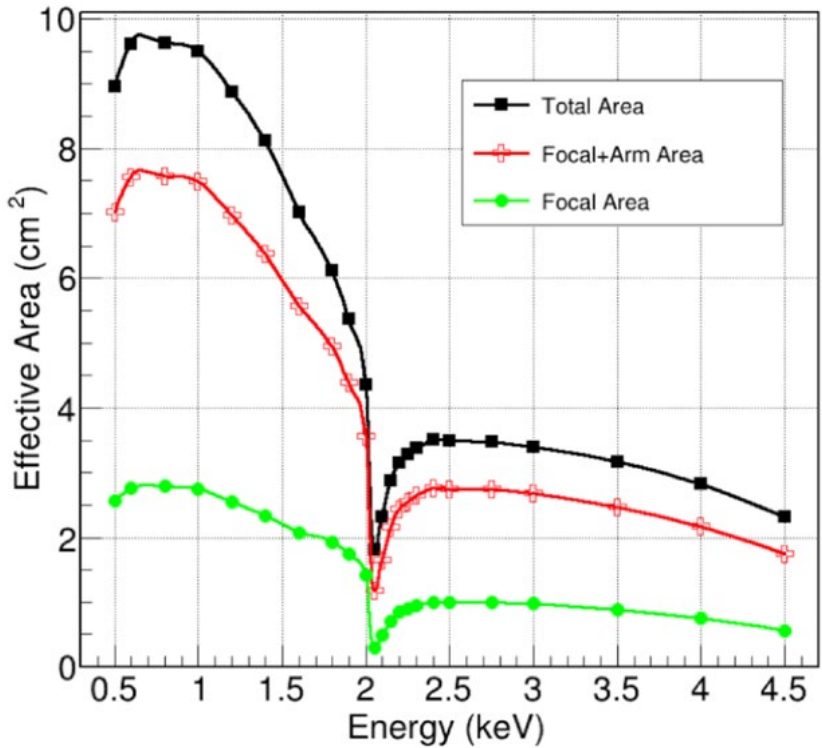

Fig. 14: Effective areas for WXT (results of simulations based on Geant4/XRTG4; see Zhao, et al. Proc. SPIE 9144, Space Telescopes and Instrumentation 2014: Ultraviolet to Gamma Ray, 91444E; doi:10.1117 /12.2055434).

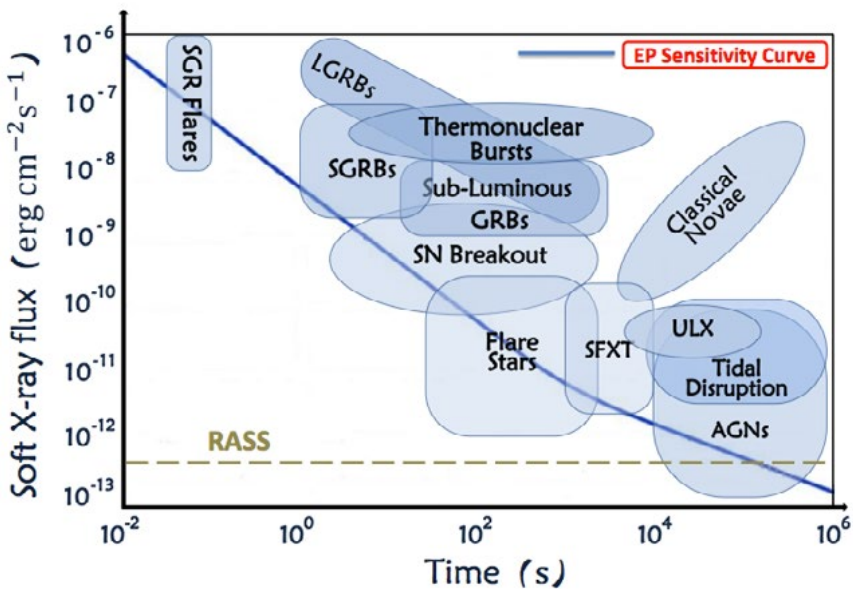

Fig. 15: Sensitivity of WXT as a function of exposure time.

The payload is of low weight (150kg) and low power consumption (200w), which can be easily accommodated within a small Chinese satellite platform. The total weight of the satellite is within 400kg including the payload based on the current design. The satellite will be in a circular orbit with an altitude of about 600 km and a period of 97 minutes, and an inclination angle about 30 degrees.

The survey strategy of Einstein Probe will be a series of pointings, each of 10-20 minutes exposure. In this way Einstein Probe will be able to cover the entire night sky (half of the sky avoiding the Sun) in about 3-5 orbits.

Bertrand Cordier introduced the French VHF network for the SVOM mission – a French-Chinese gamma-ray burst satellite, which can also provide fast downlink of the alert data of newly discovered transients. SVOM has an orbit similar to that of Einstein Probe, a circular orbit at a height of ~600 km with an inclination about 30 degrees. The requirement of SVOM for fast telemetry is to make the positional information of detected GRBs available to ground-based telescopes within 30 seconds after the trigger for 65% of the cases, and within 20 minutes for 95% of the cases. The transmission speed is 600 bits/s which is enough to downlink the basic properties of sources including position, duration, hardness ratios, etc. This network should be in operation in 2020 and could be usable by other missions, and particularly the Einstein Probe mission.

## POSSIBLE AUXILIARY INSTRUMENTS

Marco Feroci presented an interesting suggestion for the addition of a wide-field, hard X-ray monitor to the payload to broaden the detecting energy bandpass of the Einstein Probe. Although being the most sensitive optics for wide-field X-ray monitoring, the MPO lobster-eye has a limited bandpass from 0.5keV to about 5 keV, spanning only one decade in X-ray photon energy. The extension of the bandpass to higher energies of tens of keV would significantly enhance EP's X-ray spectroscopic ability, which adds valuable information for quick classification of detected transient sources and subsequent detailed diagnostic study of their physical nature. This would be particularly helpful for the cases of quick discrimination of X-ray rich GRBs from Type 1 X-ray bursts, short GRBs from magnetar short bursts, and perhaps relativistic TDEs from normal TDEs, as well as the detection of heavily absorber transients such as magnetars. For Galactic black hole candidates, the Lobster-eye monitor will detect X-ray flares (state transitions) while a hard X-ray monitor can add better constraints on the X-ray spectral parameters and their evolution with time.

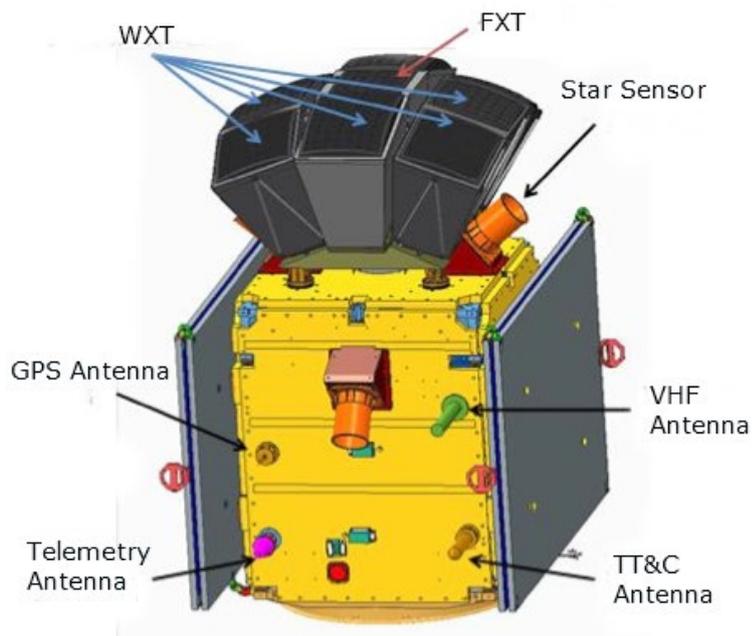

Fig. 16: Layout of the Einstein Probe satellite (courtesy of NAOC & MicroSat, CAS).

The Wide-Field X-ray Monitor (WFM) onboard LOFT – an X-ray astronomy mission proposed to ESA – is a promising candidate for such an auxiliary hard-X-ray monitor. It is a coded-mask based instrument operating in 2-50keV, which is a good matching extension to higher energies of the soft X-ray band of EP/WXT. It will provide wide-band spectra with 300eV resolution for bright X-ray sources. A set of four XRM modules would cover a large fraction of the EP/WXT field of view, at the cost of additional 50kg weight, 25W power consumption, and 40kbps telemetry rate. The LOFT/WFM has a high readiness (TRL ~ 5) and has been tested extensively.

## OTHER PROPOSED WIDE-FIELD HIGH-ENERGY ASTROPHYSICS MISSIONS

Neil Gehrels discussed a MPO-Lobster All-Sky Monitor mission concept proposed to NASA. The Lobster All-Sky Monitor is a wide-field X-ray transient detector that may be deployed either a free-flying satellite or on the International Space Station. Through its unique imaging X-ray optics that allow a 30 deg by

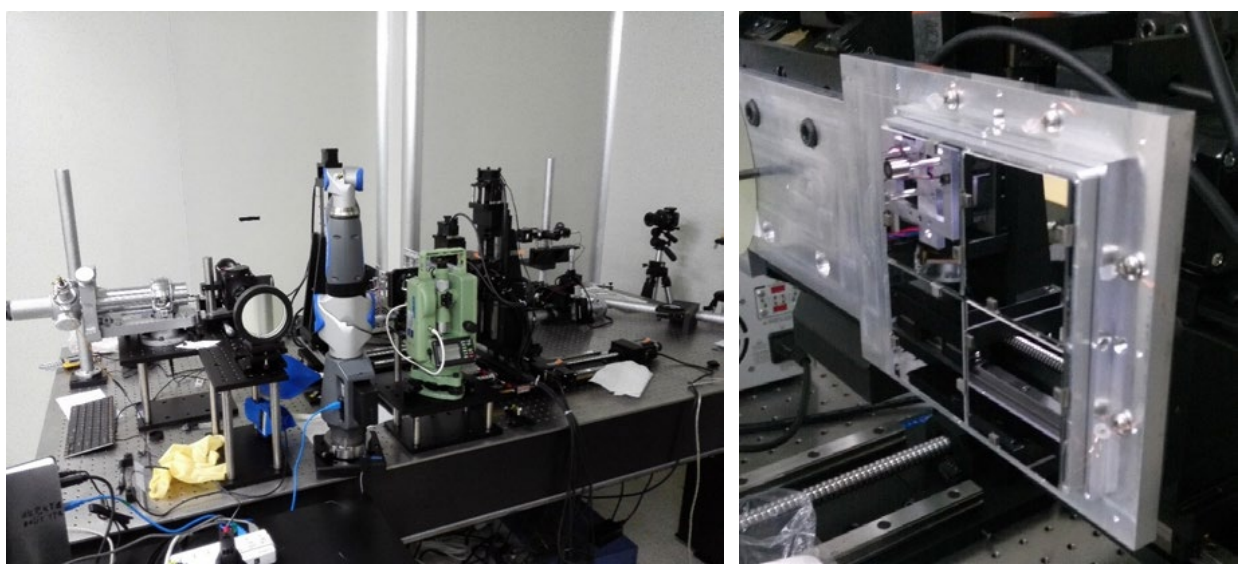

Fig. 17: Laser-guided MPO assembling and contactless online quality-monitoring system for the building of the EP/WXT. prototype module (below left). Currently an assembling precision of micro-meter or arc-second has been realized for a 2x2 MPO pieces array using a quasi-stress-free control system (below right) (credit: Chen Zhang, X-ray Imaging Lab, NAOC).

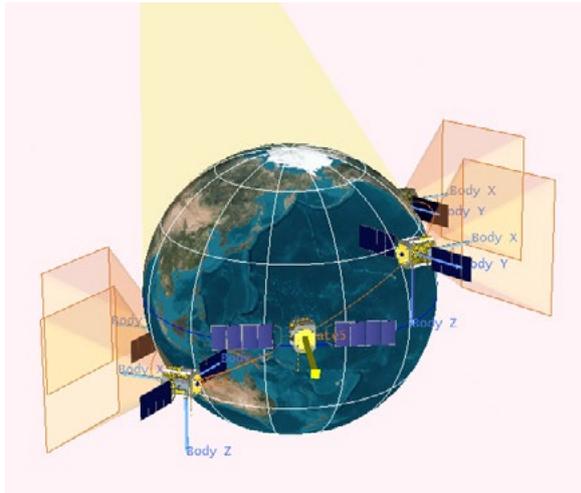

Fig. 18: Illustration of the survey mode of Einstein Probe in orbit as a series of pointed observations for 10-20 minutes exposure each, with its large instantaneous field-of-view. It will cover half of the entire sky (anti-solar direction) in 3-5 orbits (courtesy of NAOC & MicroSAT, CAS).

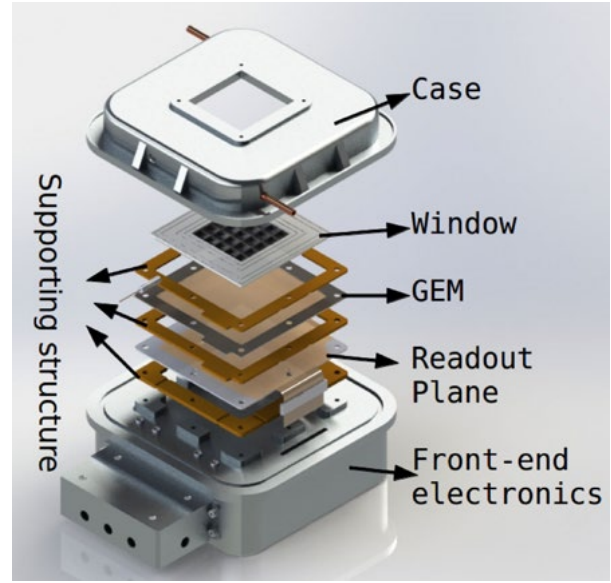

Fig. 19: Sketch of the focal-plane detector for a WXT module (credit: Hua Feng, Tsinghua University & NAOC).

30 deg FoV, a 1 arcmin position resolution and a $10^{-11}$ erg/(sec cm2) sensitivity in 2000 sec, Lobster will observe numerous events per year of X-ray transients related to compact objects, including: tidal disruptions of stars, supernova shock breakouts, neutron star bursts and superbursts, high redshift Gamma-Ray Bursts, active galactic nuclei, and perhaps most exciting, X-ray counterparts of gravitational wave detections involving both stellar mass and supermassive black holes. X-ray counterparts of gravitational waves can be localized with Lobster, allowing the determination of the host galaxy and providing the redshift and astrophysical context of the source.

Deployment of this detector on the ISS will realize cost savings compared to a free-flying satellite as power, communication, and ISS transport are provided; in this case, a 3-axis gimbal system will allow fast pointing in response to any independent, multi-wavelength indication of the above events.

## SUMMARY AND RECOMMENDATION

**1)** The Forum "Exploring the Dynamic X-ray Universe" attracted senior high-energy astrophysicists from the UK, the USA, Europe and China to Beijing, including some of the world-leading scientists in this field, to discuss the potential for new scientific discoveries provided by observing in the X-ray regime. The coming era of time-domain astronomy with large facilities across the electromagnetic spectrum, and in gravitational waves and neutrinos, demands a matching high-energy observational capability.

**2)** Transient high-energy source discovery has been a major driver of scientific advance continuously since the early days of X-ray astronomy. Many new types of celestial objects have been found by the wide-field high-energy monitors. The US/UK/Italian satellite Swift, launched in 2004, is highly prolific, combining a Burst Alert Telescope, able to view one sixth of the sky at any time, with narrow-field X-ray and UV/optical follow-up telescopes on an autonomous and rapidly slewing spacecraft.

**3)** The end of this decade will see the dawn of gravitational wave astronomy, with the direct detection of gravitational waves by the advanced gravitational wave detectors such as Advanced LIGO and VIRGO. The scientific importance of this widely anticipated result is hard to overstate, but the simultaneous detection of their electromagnetic counterparts is essential to place them in a familiar astrophysical context so that the maximum information can be extracted from these complementary channels. The X-ray band is ideal for the prompt detection of these transients.

**4)** Wide-field X-ray instruments have suffered from poor sensitivity due to the need to use collimation to provide spatial resolution. A new technology is now becoming available which will revolutionise these instruments: focusing wide-field X-ray optics. Following the geometry of the eye of a lobster, the micro-pore X-ray optic can achieve an arbitrarily large field of view. The MIXS instrument on ESA's BepiColombo mission to Mercury, due for launch in 2016,

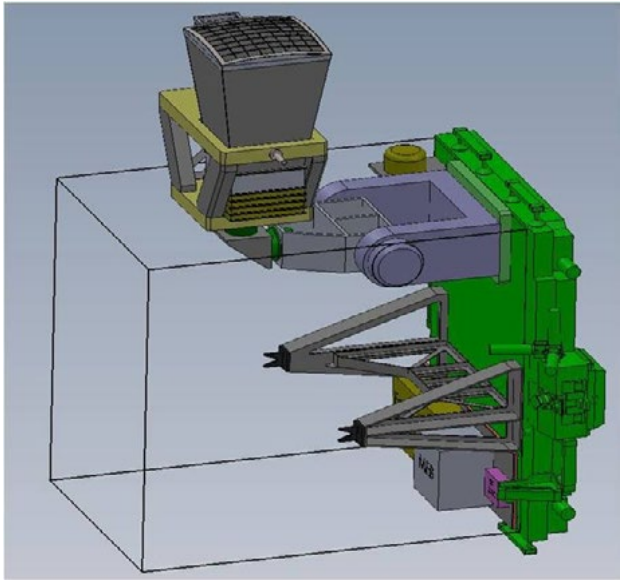

Fig. 20: ISS-Lobster showing pointing platform (credit: N. Gehrels & J. Camp, NASA).

will be the first to fly with micro-pore optics for X-ray imaging.

**5)** The thousand-year-long Chinese quest for the heavens is manifested today in a series of ambitious national science programmes, with satellite-borne facilities a major focus, featuring Einstein Probe and the others. The Chinese orbiting wide-field high-energy observatory concept, Einstein Probe, was described in detail. This will be the first to take advantage of micro-pore lobster optics, and has a very promising discovery potential. Consisting of a 60°x60° soft X-ray set of wide-field lobster telescopes, together with a sensitive narrow-field follow-up X-ray telescope on a robotic and rapidly slewing spacecraft with prompt downlink capability, Einstein Probe would have unprecedented sensitivity, and would be able to trigger follow-up observations of a wide range of source types from facilities around the world.

**6)** The forum participants enthusiastically endorsed the Einstein Probe mission concept as it promises unique and truly cutting-edge science, which would revolutionise time-domain high-energy astrophysics, and make China a world leader in this exciting field.

**7)** The combination of very wide field of view X-ray telescopes with a more sensitive follow-up X-ray telescope automatically and quickly re-pointed was seen as providing a very strong and self-contained science capability.

**8)** The forum recommended that the Einstein Probe team proceed as quickly as possible to implementation of the mission within the Chinese programme, taking advantage the University of Leicester's unique expertise in micro-pore Lobster optics.

**9)** The scientists present offered continued support in the further elaboration of the scientific case for Einstein Probe and in the development of some of the key technologies, and to foster relations with other world-class astrophysics facilities to maximize the scientific impact of this exciting mission.

## ACKNOWLEDGE-MENT

We are very grateful to all of the participants. Special thanks are given to Dr. Maurizio Falanga, Executive Director of ISSI-Beijing, for his great efforts and high efficiency in organising this successful meeting, as well as Lijuan En (ISSI-Beijing), Ariane Bonnet (formerly ISSI-Beijing) and Congying Bao (NAOC) for their wonderful work to make everything proceed in a professional, smooth and comfortable way. We would also like to thank Sabrina Brezger (ISSI-BJ) for the editing this report. We are grateful to ISSI-Beijing and NSSC (especially the Director, Prof. Ji Wu) for their kind hospitality during the forum.
Furthermore, we are grateful to the following participants and colleagues for kindly providing some of the contents and material of this report: R. Willingale, N. Gehrels, J. Camp, C. Zhang, H. Feng.

W. Yuan & J. P. Osborne

***Note added***: In May 2015 Einstein Probe was selected as a 'candidate mission of priority' (mission would go ahead without further selection once the funding is secured for the space science programme of the CAS), with a proposed launch date around 2020.

# TAIKONG

## Participants

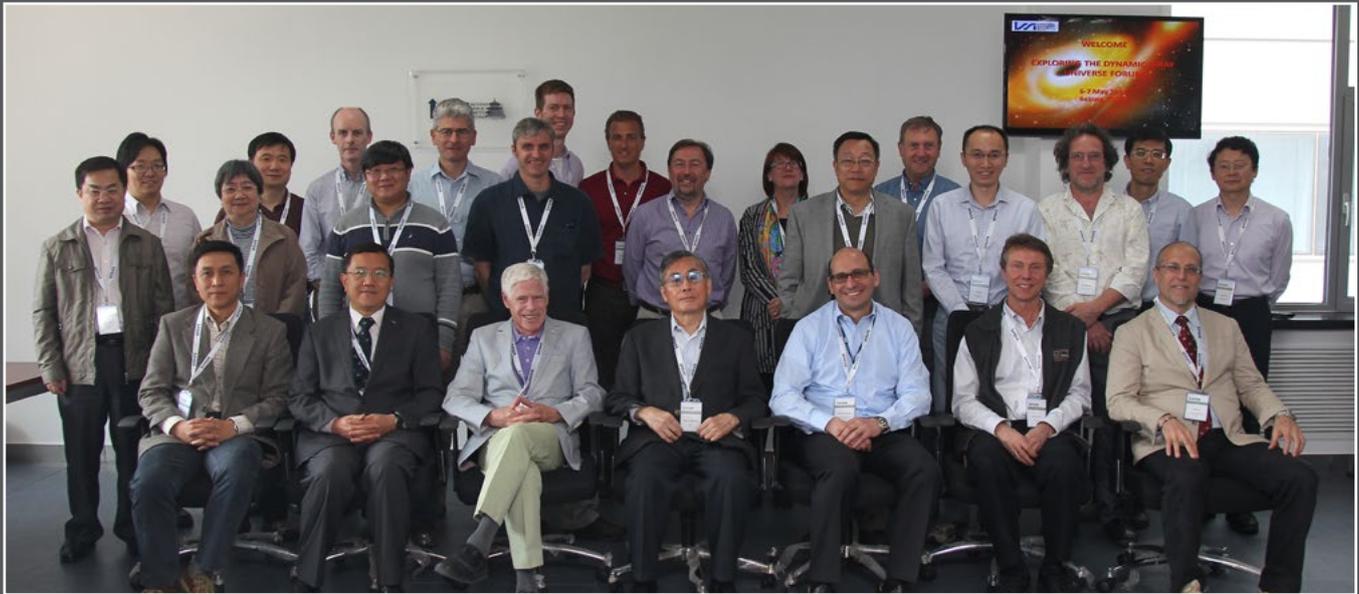

| | |
|---|---|
| Roger M. Bonnet | International Space Science Institute, Switzerland |
| Phil Charles | Southampton University, UK |
| Bertrand Cordier | CEA, France |
| Zigao Dai | Nanjing University, China |
| Maurizio Falanga | International Space Science Institute - Beijing, China |
| Hua Feng | Tsinghua University, China |
| Marco Feroci | INAF-Rome, Italy |
| Neil Gehrels | NASA/GSFC, USA |
| Lijun Gou | National Astronomical Observatories, CAS, China |
| Stefanie Komossa | Max-Planck Institute for Radio Astronomy, Germany |
| Jon Lapington | Leicester University, UK |
| Xiangdong Li | Nanjing University, China |
| Fukun Liu | Peking University, China |
| Yuqian Ma | Institute of High-Energy Physics, CAS, China |
| Masaru Matsuoka | RIKEN, Japan |
| Paul O'Brien | Leicester University, UK |
| Julian Osborne | Leicester University, UK |
| Luigi Piro | INAF-Rome, Italy |
| Patrick Sutton | Cardiff University, UK |
| Richard Willingale | Leicester University, UK |
| Ji Wu | National Space Science Center, CAS, China |
| Xuefeng Wu | Purple Mountain Observatory, CAS, China |
| Weimin Yuan | National Astronomical Observatories, CAS, China |
| Bing Zhang | University of Nevada, USA |
| Chen Zhang | National Astronomical Observatories, CAS, China |
| Shuang-Nan Zhang | Institute of High-Energy Physics, CAS, China |